\documentclass[final,5p,times,twocolumn]{elsarticle}

\usepackage{graphicx}
\usepackage{amssymb}
\usepackage{amsmath}
\usepackage{threeparttable}

\journal{Nuclear Instruments and Methods in Physics Research A}

\begin{document}

\begin{frontmatter}

%% Title, authors and addresses

%% use the tnoteref command within \title for footnotes;
%% use the tnotetext command for the associated footnote;
%% use the fnref command within \author or \address for footnotes;
%% use the fntext command for the associated footnote;
%% use the corref command within \author for corresponding author footnotes;
%% use the cortext command for the associated footnote;
%% use the ead command for the email address,
%% and the form \ead[url] for the home page:
%%
%% \title{Title\tnoteref{label1}}
%% \tnotetext[label1]{}
%% \author{Name\corref{cor1}\fnref{label2}}
%% \ead{email address}
%% \ead[url]{home page}
%% \fntext[label2]{}
%% \cortext[cor1]{}
%% \address{Address\fnref{label3}}
%% \fntext[label3]{}

\title{Spatial resolution of a $\mu$PIC-based neutron imaging detector}

%% use optional labels to link authors explicitly to addresses:
%% \author[label1,label2]{<author name>}
%% \address[label1]{<address>}
%% \address[label2]{<address>}

\author[ku]{J.D.~Parker\corref{cor1}}
\ead{jparker@cr.scphys.kyoto-u.ac.jp}

\author[jaea]{M.~Harada}
\author[ku]{K.~Hattori}
\author[ku]{S.~Iwaki}
\author[ku]{S.~Kabuki}
\author[ku]{Y.~Kishimoto}
\author[ku]{H.~Kubo}
\author[ku]{S.~Kurosawa}
\author[ku]{Y.~Matsuoka}
\author[ku]{K.~Miuchi}
\author[ku]{T.~Mizumoto}
\author[ku]{H.~Nishimura}
\author[jaea]{T.~Oku}
\author[ku]{T.~Sawano}
\author[jaea]{T.~Shinohara}
\author[jaea]{J.~Suzuki}
\author[ku]{A.~Takada}
\author[ku]{T.~Tanimori}
\author[ku]{K.~Ueno}

\cortext[cor1]{Corresponding author}

\address[ku]{Department of Physics, Graduate School of Science, Kyoto University, Kitashirakawa-oiwakecho, Sakyo-ku, Kyoto 606-8502, Japan}
\address[jaea]{Materials and Life Science Facility Division, Japan Atomic Energy Agency (JAEA), Tokai, Ibaraki 319-1195, Japan}

\begin{abstract}
%% Text of abstract

We present a detailed study of the spatial resolution of our time-resolved
neutron imaging detector utilizing a new neutron position reconstruction method 
that improves
both spatial resolution and event reconstruction efficiency.
Our prototype detector system, employing a micro-pattern gaseous
detector known as the micro-pixel chamber ($\mu$PIC) coupled with a 
field-programmable-gate-array-based data acquisition system,
combines 100$\mu$m-level spatial and sub-$\mu$s time resolutions
with excellent gamma rejection and high data rates, 
making it well suited for applications in neutron radiography
at high-intensity, pulsed neutron sources.
From data taken at the Materials and Life Science Experimental Facility
within the Japan Proton Accelerator Research Complex (J-PARC),
the spatial resolution 
was found to be approximately Gaussian with a sigma of 
$103.48 \pm 0.77$~$\mu$m (after correcting for beam divergence).
This is a significant improvement over that achievable with our previous
reconstruction method ($334 \pm 13$~$\mu$m),
and compares well with conventional neutron imaging detectors
and with other high-rate detectors currently under development.
Further, a detector simulation indicates that a spatial resolution of less than
60~$\mu$m may be possible with optimization of the gas characteristics 
and $\mu$PIC structure.
We also present an example of imaging combined with 
neutron resonance absorption spectroscopy.

\end{abstract}

\begin{keyword}
%% keywords here, in the form: keyword \sep keyword
neutron imaging \sep gaseous detector \sep micro-pattern detector

%% MSC codes here, in the form: \MSC code \sep code
%% or \MSC[2008] code \sep code (2000 is the default)

\end{keyword}

\end{frontmatter}

%%
%% Start line numbering here if you want
%%
% \linenumbers

%% main text
\section{Introduction}
\label{sec:intro}

We have developed a time-resolved neutron imaging detector
with sub-mm spatial and $\mu$s-order time resolutions, excellent gamma
rejection, and high-rate capability~\cite{parker12}.
This neutron imaging detector is intended for use at high-intensity, 
pulsed neutron sources,
where radiographic imaging is combined with neutron energy measured
via time-of-flight (TOF).
Our detector consists of a time-projection-chamber (TPC) with an
active volume of $10 \times 10 \times 2.5$~cm$^3$ contained within an aluminum
pressure vessel.
The TPC is read out by a $\mu$PIC (micro-pixel chamber) 
micro-pattern gaseous detector with a 400-$\mu$m pitch, incorporating
orthogonal anode and cathode strips for a 2-dimensional readout~\cite{ochi01}.
Using an Ar-C$_2$H$_6$-$^3$He (63:7:30) gas mixture at 2 atm, 
neutrons are detected via absorption on $^3$He
with 18\% efficiency at a neutron energy of 25.3~meV. 
A peak efficiency of up to 65\% at 0.35~meV is possible with the simple 
modifications to the pressure vessel described in Ref.~\cite{parker12}.
The detector is coupled with a fast, 
FPGA (Field Programmable Gate Array)-based data acquisition 
system~\cite{orito04,kubo05}
that records both the energy deposition (via time-over-threshold) 
and 3-dimensional tracking information (2-dimensional position plus time)
for each neutron event.

Our neutron imaging detector is well suited to such TOF-based techniques as
neutron resonance absorption spectroscopy (NRAS)
for measuring nuclide composition, density, and temperature~\cite{res:sato09},
and Bragg-edge transmission for studying
crystal structure (e.g., lattice spacing, grain size and 
orientation, etc.) and residual strain~\cite{bragg:sato09,bragg:santis02}.
Such an area detector enables detailed, non-destructive
2-dimensional and 3-dimensional (using computed tomography) study of internal 
material properties at the sub-mm level with short measurement times.

As reported in Ref.~\cite{parker12},
our detector achieved a spatial resolution of $349 \pm 43$~$\mu$m ($\sigma$), 
a time resolution as low as 0.6~$\mu$s,
and an effective gamma sensitivity of 10$^{-12}$ or less.
In the present paper,
we introduce an analysis method that more fully leverages the 
time-over-threshold information to significantly improve the accuracy and
efficiency of the neutron position reconstruction.
We then present a study of the uniformity and gain dependence of the 
spatial resolution,
discuss the results in relation to other neutron imaging detectors,
and end with a simple demonstration of radiographic and resonance imaging.

\section{Neutron position reconstruction}
\label{sec:recon}

The absorption of low energy neutrons on $^3$He
produces a proton-triton pair with total kinetic energy of $\sim$764~keV
and a combined track length of about 8~mm in the 2-atm gas of our detector.
As the range of the proton is about three times that of the heavier triton,
the separation of the two particles is essential for an accurate
determination of the neutron interaction position.
The separation is complicated, however, by the fact that, 
due to the low incident neutron energies typical in radiography, 
the proton and triton emerge essentially back-to-back.
It is thus necessary to utilize additional information, 
in the form of the energy deposition at each strip
(estimated via time-over-threshold), to perform the separation.

\begin{figure}[ht]
\centering
\includegraphics[width=7. cm,clip]{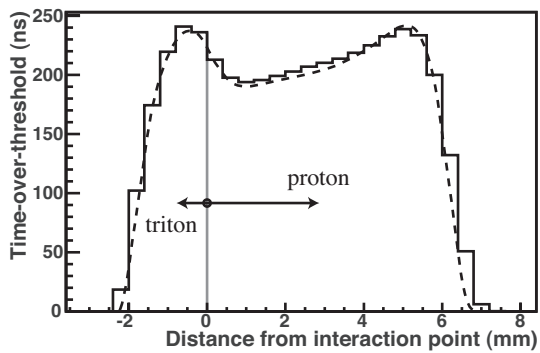}
\caption{\label{fig:tot} Time-over-threshold (TOT) for proton-triton tracks.
This representative TOT distribution (solid line) was determined by
averaging several thousand neutron events with a particular proton-triton 
track angle.
The simulation result for the same track angle is overlaid (dashed line).
Arrows indicate the direction of the proton and triton.
}
\end{figure}

In Ref.~\cite{parker12}, the shape of the energy deposition,
like that illustrated in Fig.~\ref{fig:tot},
was used to determine the proton direction,
and a correction to the neutron position was then made
relative to the mid-point of the track.
This simple algorithm achieved a spatial resolution of 
$349 \pm 43$~$\mu$m ($\sigma$),
representing nearly a factor of three improvement over that possible in the absence
of the time-over-threshold (TOT) information 
(i.e., when taking the neutron position as the mid-point of the 
proton-triton track).
However,
the algorithm suffered from poor reconstruction efficiency, since it
was only applicable when the proton and triton were clearly visible
in both the anode and cathode TOT distributions.
This requirement severely restricted the usable range of track angles,
resulting in the rejection of about 65\% of detected neutron events.

In the present paper,
we introduce a more accurate method for finding the neutron position that
not only produced a better spatial resolution but allowed us to include
the entire range of track angles.
Here,
the neutron position was found by fitting the measured TOT distributions
with the expected distributions (referred to as {\em templates}) 
determined using a simulation of our detector system.
The simulation was based on the GEANT4 software toolkit~\cite{geant4}
as described in Ref.~\cite{parker12}.
The simulation result for a particular proton-triton track angle is shown in 
Fig.~\ref{fig:tot} (dashed line) along
with the TOT distribution for real neutrons (solid line) at the same angle.
The slight disagreement between the distributions is due to simplifications
and uncertainties in the simulated response of the data acquisition hardware,
coupled with the uncertainty in the neutron interaction position in the 
real data.
The fitting, along with all data analysis for the results presented here, 
was carried out using custom software based on the ROOT object-oriented 
framework~\cite{root}.

\section{Spatial resolution}
\label{sec:res}

The spatial resolution was studied using
images of cadmium test charts taken in March 2012 
at NOBORU (NeutrOn Beamline for Observation and Research Use),
located at Beamline 10 of the Materials and Life Science Experimental Facility (MLF)
within the Japan Proton Accelerator Research Complex (J-PARC)~\cite{noboru09}.
The neutron pulse rate was 25~Hz 
(for a total neutron band-width at NOBORU of 10~\AA) 
with a beam power of 200~kW, and
a $3.2 \times 3.2$~mm$^2$ collimator near the midpoint of the beamline
provided a low-dispersion beam suitable for high-resolution imaging.
The detector, located 15~m from the moderator,
was filled with Ar-C$_2$H$_6$-$^3$He (63:7:30) at 2~atm
and equipped with a 2.5-cm drift cage as in Ref.~\cite{parker12},
and the test charts
were placed directly on the entrance window of the detector vessel.
The detector was operated at a gas gain of 470, unless indicated otherwise.

\begin{figure}[ht]
\centering
\includegraphics[width=7.5 cm,clip]{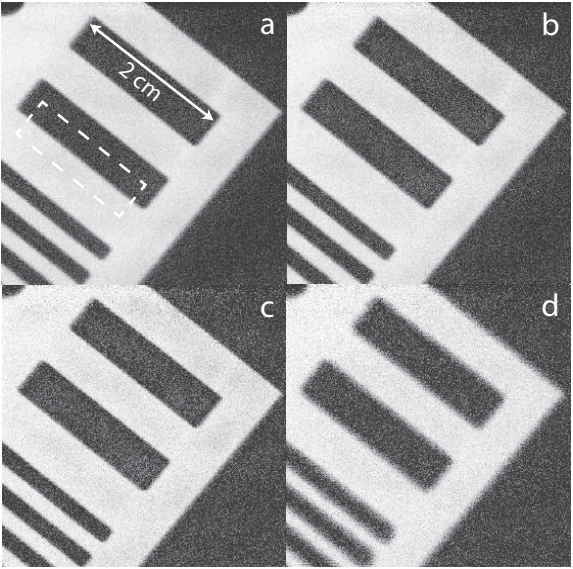}
\caption{\label{fig:edge45} Images of a Cd test chart.
Images reconstructed using the template method for
(a) cut case 1, (b) cut case 2, and (c) cut case 3 as defined in
the text.
The image in (d) was reconstructed from the same data set
using the simple algorithm of Ref.~\cite{parker12} and the same cuts as (c).
Darker regions indicate higher neutron transmission.
Bin size is 100~$\mu$m.
}
\end{figure}

Under these conditions,
data was taken of one test chart for an exposure time of
7.8 minutes (not including data read-out time),
and the resulting images, normalized to the beam profile, 
are shown in Fig.~\ref{fig:edge45} for the various cuts described below.
In the offline analysis,
neutrons with energy well below the cadmium cut-off were selected
by requiring a TOF greater than 2.1~ms ($270 > E_{\rm neutron} > 0.74$~meV, 
with the lower limit arising from the 10~\AA~bandwidth of NOBORU),
and the gamma rejection cuts on total energy deposit and 
proton-triton track length described in our previous work were applied.
Images were then reconstructed for three cut cases:
1) the above TOF and gamma cuts,
2) an additional consistency cut on the proton direction, and
3) the further requirement that the anode and cathode TOT distributions
each displayed two distinct peaks (as in Fig.~\ref{fig:tot}).
The consistency cut required agreement in the component of the proton direction
perpendicular to the plane of the $\mu$PIC (determined using the hit times)
as measured independently by the anodes and cathodes.
For cut case 1, the inconsistent events were {\em corrected} by changing the
proton direction for the component with fewer hits, while
they were simply removed in the second case.
The last cut is the proton-triton separation (PTS) cut used in the 
method of Ref.~\cite{parker12}.
The images of Figs.~\ref{fig:edge45}(a), \ref{fig:edge45}(b), 
and  \ref{fig:edge45}(c)
correspond to cut cases 1, 2, and 3, respectively.
The image of Fig.~\ref{fig:edge45}(d), reconstructed using the simple method
of Ref.~\cite{parker12} with the same cuts as \ref{fig:edge45}(c), 
is shown for comparison.
The number of events remaining after each cut 
are listed in Table~\ref{tbl:reff},
along with the event reconstruction efficiency, $\varepsilon_{rec}$, defined
as the fraction of usable neutron events (i.e., those passing the TOF and 
gamma rejection cuts) included in the final images.

\begin{table}[ht]
\caption{\label{tbl:reff} Image reconstruction efficiency.
Listed here are the number of events remaining after various selection cuts,
along with the event reconstruction efficiency, $\varepsilon_{rec}$,
for the imaging of a Cd test chart.
}
\begin{center}
\tabcolsep=0.15cm
\small
\begin{tabular}{lllll}
\hline
Cut & Number & Fraction & $\varepsilon_{rec}$ (\%) & Figure \\
\hline
%Beam pulses                      & 11756               & $-$   & \\ 
No cut                           & $4.20 \times 10^{7}$ & $-$   & & \\ 
TOF $> 2.1$~ms                   & $3.42 \times 10^{7}$ & 1     & & \\ 
Energy $> 60$~clocks             & $3.37 \times 10^{7}$ & 0.986 & & \\ 
Track length ($-2$/$+$3$\sigma$) & $2.77 \times 10^{7}$ & 0.812 & 100 & 2(a) \\ 
Consistency cut                  & $1.83 \times 10^{7}$ & 0.536 & 66.0 & 2(b) \\ 
PTS cut                          & $8.83 \times 10^{6}$ & 0.258 & 31.8 & 2(c,d) \\ 
\hline
\end{tabular}
\end{center}
\end{table}

\begin{figure}[ht]
\centering
\includegraphics[width=7.5 cm,clip]{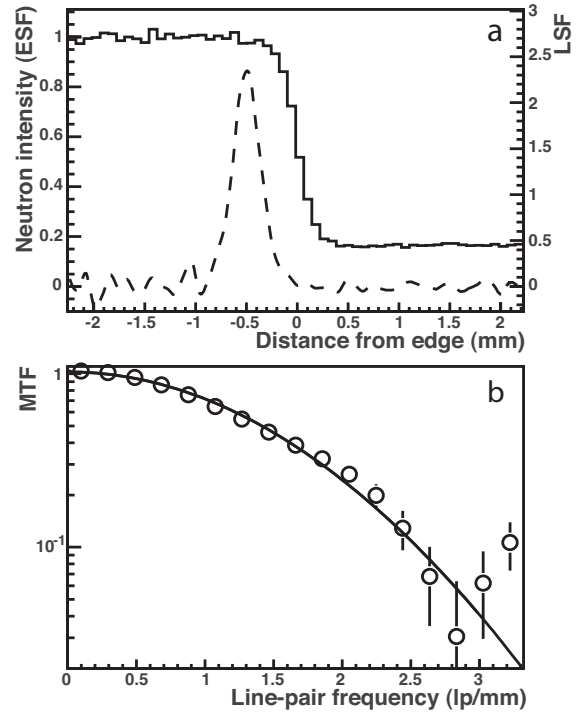}
\caption{\label{fig:mtf} The modulation transfer function.
Plot (a) shows the ESF (solid line) and
LSF (dashed line) for the slit edge indicated in Fig.~\ref{fig:edge45}(a).
The LSF has been shifted by $-0.5$~mm for ease of viewing.
Bin size is 80~$\mu$m.
Plot (b) shows the modulation transfer function (MTF) overlaid with
the best-fit Gaussian.
}
\end{figure}

A detailed study of the spatial resolution was carried out
using a 1.4-cm edge-section oriented at an angle of about 40$^{\circ}$ relative 
to the anode strip direction 
(indicated in Fig.~\ref{fig:edge45}(a) by the dotted line).
The resolution was studied using the modulation transfer function (MTF),
calculated via Fourier transform from the discrete derivative of 
the intensity distribution across the edge, as in Ref.~\cite{hattori12}.
The observed edge spread function (ESF) 
and its derivative (line spread function, or LSF) 
are shown in Fig.~\ref{fig:mtf}(a) for cut case 1,
with the resulting MTF shown in Fig.~\ref{fig:mtf}(b).
(The distributions for the remaining cases were similar.)
The MTF displays a clear Gaussian character and drops to 10\% at around
2.5 lp/mm (line pairs/mm), or a line width of $\sim$200~$\mu$m.
The Gaussian shape indicates that the correction for events
with inconsistent proton direction is effective,
since any significant contribution from events with incorrect proton direction
would result in long tails in the ESF and LSF,
producing a Lorentzian-like resolution function with an exponential MTF.
From the shape of the ESF of Fig.~\ref{fig:mtf}(a)
(or more specifically, its deviation from an error function), 
we estimate any remaining contamination from events with incorrect proton direction 
at less than 5\%,
with the effect on the measured resolution for cut case 1 smaller than
our current error level.
The values of the spatial resolution ($\sigma_{res}$) 
for Figs.~\ref{fig:edge45}(a), (b), and (c),
determined by a Gaussian fit to the MTF,
are listed in Table~\ref{tbl:mtf}.
Also listed are the resolutions determined from a second set of images 
with the chart rotated to orient the edge at roughly 0$^{\circ}$ 
(i.e., nearly parallel to the anode strip direction),
along with the weighted averages.
The small differences in the results for the two edges are due mainly to their
different positioning relative to the $\mu$PIC and beam center.
The spatial resolutions for images reconstructed using the 
previous method [as in Fig.~\ref{fig:edge45}(d)] are also listed, 
and were found to be in good agreement with the result 
of our previous paper ($349 \pm 43$~$\mu$m~\cite{parker12}).

\begin{table}[ht]
\caption{\label{tbl:mtf} Spatial resolution from a Cd chart.
Spatial resolutions ($\sigma_{res}$) for the three different cut conditions
described in
the text, as determined using edges oriented at 40$^{\circ}$ and 0$^{\circ}$
relative to the horizontal.
Also listed are the resolutions for the same images reconstructed using the
simple method of Ref.~\cite{parker12}.
All values are given in $\mu$m.
}
\begin{center}
\begin{tabular}{lccc}
\hline
Cut & \multicolumn{3}{c}{$\sigma_{res}$} \\ \cline{2-4}
    & 40$^{\circ}$ edge & 0$^{\circ}$ edge & Average \\
\hline
Tracking & $130.7 \pm 3.1$ & $136.5 \pm 2.8$ & $133.9 \pm 2.1$ \\
Consistency & $125.0 \pm 3.3$ & $131.5 \pm 3.1$ & $128.5 \pm 2.3$ \\
PTS & $111.2 \pm 4.0$ & $114.4 \pm 3.7$ & $112.9 \pm 2.7$ \\
\hline 
Old method & $325 \pm 19$ & $343 \pm 18$ & $334 \pm 13$ \\
\hline
\end{tabular}
\end{center}
\end{table}

The improvement in the resolution observed between the three cut cases
can be understood as
the result of the dependence of the spatial resolution 
and proton direction determination on the proton-triton track angle,
each worsening as the angle of the track approaches the
perpendicular to the plane of the $\mu$PIC.
Such tracks will have a smaller two-dimensional projection 
(i.e., the charge will be deposited in fewer strips),
resulting in poorer fit results due to fewer data points
and increasing ambiguity of the proton direction due to
overlap of the proton and triton Bragg peaks.
The consistency cut (cut case 2) thus tends to remove these ambiguous,
low-information tracks,
resulting in improved spatial resolution.
Cut case 3 restricts the angles further by preferring tracks
that are more parallel to the $\mu$PIC and distributed around 45$^\circ$
relative to the anode and cathode strip directions,
essentially maximizing the quality of the tracks and 
achieving the best spatial resolution.

\subsection{Uniformity of the spatial resolution}
\label{sec:unif}

\begin{figure}[ht]
\centering
\includegraphics[width=6. cm,clip]{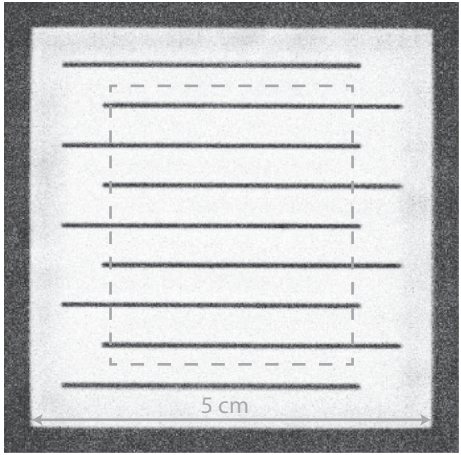}
\caption{\label{fig:slits} Image of slit pattern.
A 0.5-mm thick Cd test chart with a pattern of 0.5-mm wide slits 
was used to study the uniformity of the spatial resolution.
The gray dashed line indicates the area used for the study.
Darker areas indicate higher neutron transmission.  Bin size is 100~$\mu$m.
}
\end{figure}

To study the uniformity of the spatial resolution,
an image of a regular slit pattern,
shown in Fig.~\ref{fig:slits}, was taken under the experimental conditions
described previously with an effective exposure of 9.4 minutes.
The slit pattern consisted of 0.5-mm wide slits cut into a 0.5-mm thick Cd
plate at a pitch of 5~mm,
and the chart was positioned with slits oriented along the anode strip direction.
The resolution at each slit was determined by fitting the function:

\begin{equation}
N(x) 
 = \frac{1}{2} A \left[ {\rm erf}\left( \frac{x - (\mu - a)}{\sqrt{2} \sigma} 
\right) - {\rm erf}\left( \frac{x - (\mu + a)}{\sqrt{2} \sigma} \right) \right] + C
\end{equation}
derived from the convolution of a Gaussian resolution with 
two step functions representing the slit.
In the above fit function, 
$A$ is a normalization factor, 
erf is the usual Gaussian error function,
$\mu$ and $a$ are the mean position and half-width of the slit, respectively,
$\sigma$ is the Gaussian resolution,
and $C$ accounts for the transmission of the Cd chart.

\begin{figure}[ht]
\centering
\includegraphics[width=7. cm,clip]{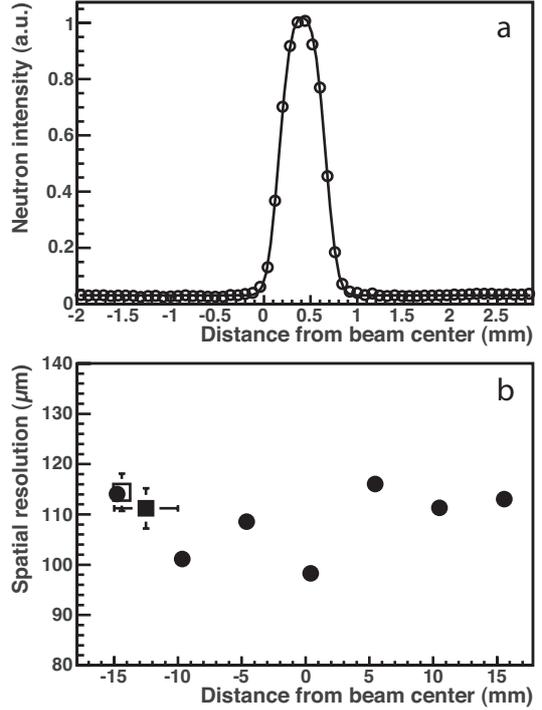}
\caption{\label{fig:ures} Uniformity of spatial resolution from slit pattern.
Figure (a) shows the fit to the intensity distribution of a single 0.5-mm slit.
Figure (b) shows the spatial resolutions determined for seven slits spaced at 
5-mm intervals (circles), along with the resolutions determined via the MTF 
from the 40$^{\circ}$ (filled square) and 0$^{\circ}$ (open square) edges.
}
\end{figure}

Fig.~\ref{fig:ures}(a) shows a fit to one slit using the
above equation, and 
Fig.~\ref{fig:ures}(b) shows the spatial resolution found from seven such
slits (circles) covering an area of roughly $3 \times 3$~cm$^2$ near the
beam center. 
The resolution determined from the slits is in good agreement with that
determined previously from the edges using the MTF method,
indicated in Fig.~\ref{fig:ures}(b) by the filled square (40$^{\circ}$ edge) 
and open square (0$^{\circ}$ edge).
The weighted average of all measurements yields a spatial resolution of 
$108.24 \pm 0.74$~$\mu$m with a root-mean-square deviation (RMSD) of 6.4\%,
indicated as a percent of the spatial resolution.

The observed non-uniformity in the spatial resolution,
represented by the RMSD, 
arises from a combination of effects
including parallax, variations in the $\mu$PIC gain and 
channel-by-channel amplifier responses,
the underlying strip structure, and statistical fluctuations.
As described in Ref.~\cite{parker12},
parallax results from beam divergence coupled with the gas depth 
of the detector.
Using our GEANT4 simulation with the beam geometry included,
the worsening of the resolution due to the parallax effect was estimated 
at 2\% near the beam center, increasing to 8\% near $\pm$1.5~cm.
Correcting the measured resolutions based on the simulation results 
gives a spatial resolution of 
$103.48 \pm 0.77$~$\mu$m with an improved RMSD of 5.3\%.
Our GEANT4 simulation also indicates that
the variations in the $\mu$PIC gain and channel-by-channel amplifier responses
contribute about 15\% to the remaining non-uniformity.
Eliminating these variations, however, also results in a corresponding
improvement in the spatial resolution such that the RMSD, as a fraction
of the spatial resolution, remains roughly constant.
Finally, while some small improvement can be expected with increased 
statistics, the effect of the underlying strip structure is difficult to
address directly.

\subsection{Spatial resolution versus gain}
\label{sec:resvg}

\begin{figure}[ht]
\centering
\includegraphics[width=7. cm,clip]{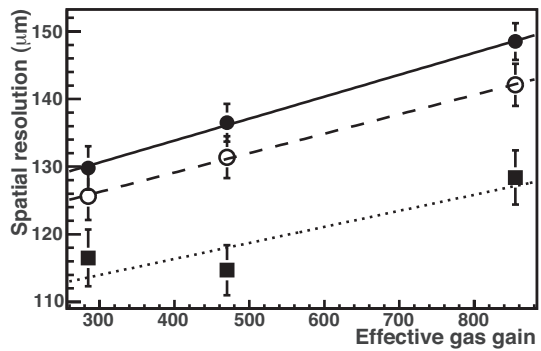}
\caption{\label{fig:gain} Spatial resolution versus gas gain.
The results for the spatial resolution are plotted versus gas gain for three 
cut conditions: tracking cuts only (solid circles), 
with consistency cut (open circles),
and with PTS cut (solid squares).
Best fit lines are overlaid for each case.
}
\end{figure}

To check the effect of the gas gain on the spatial resolution,
images of the edge at 0$^{\circ}$ were taken at two additional gain settings,
one higher ($\sim$850) and one lower ($\sim$285) than that used above,
while the thresholds that determine TOT were held constant.
The images were reconstructed using templates optimized for each gain setting, and
the resulting spatial resolutions, found using the MTF procedure, are 
plotted in Fig.~\ref{fig:gain} for each of the three cut conditions.
The resolution shows a clear trend, 
improving by roughly 3~$\mu$m for each decrease in gain of 100.
A similar trend was observed for images generated with our GEANT4 simulation
and was seen to continue to lower gain settings.
This improvement can be understood as a result of the non-linearity of
the relationship between pulse height and TOT~\cite{hattori12},
tending toward saturation for larger pulse heights (i.e., higher gains) 
and thereby decreasing the prominence 
of the proton and triton peaks in the TOT distribution.

This result indicates that a lower gas gain, or equivalently, that a smaller
pulse-height-to-threshold ratio,
is preferred for neutron imaging with our detector.
The thresholds used for all measurements presented here 
(anodes: $-$2.5~mV, cathodes: $+$3.6~mV, 
with the difference arising from a larger intrinsic amplifier gain 
for positive pulses from the cathodes as discussed in Ref.~\cite{hattori12}) 
were determined as the minimum needed to reject essentially all digital noise.
While it may be possible to improve the spatial resolution at higher
gain settings by increasing the discriminator thresholds,
using the lowest possible gain (and thus, lowest thresholds) 
has two important advantages:
1) a low gain requirement increases the usable life of the gas 
for minimum $^3$He consumption, and 
2) the electrical current generated on the $\mu$PIC is minimized, 
by minimizing the charge deposition per event, for stable high-rate operation.
The lowest gain is limited by the neutron detection efficiency,
which, while constant at higher gains,
drops off rapidly below a gain of $\sim$200.
This suggests a practical gain setting of around 250 for neutron imaging,
safely within the region of constant efficiency.
This would give an additional improvement of 5 to 6~$\mu$m in the 
spatial resolutions determined in the previous section.
Incidentally,
these optimized gain and threshold settings coincide with those shown 
in Ref.~\cite{parker12} to give a gamma sensitivity of $<$10$^{-12}$.

\section{Discussion}
\label{sec:disc}

The results of the previous section show that
our $\mu$PIC-based detector can achieve a spatial resolution of
just under 100~$\mu$m for the tightest cut condition 
(or $\sim$120~$\mu$m for all usable neutron events)
after reducing the gas gain to 
the optimum value of 250 and correcting for beam divergence.
This resolution compares well with that of imaging plates 
(25 to 50~$\mu$m)~\cite{nip_kobayashi99}
and typical CCD detectors ($\sim$100$\mu$m)~\cite{ccd_lehmann07},
with the advantages over these conventional imaging detectors
of sub-$\mu$s time resolution and excellent gamma discrimination
as discussed in Ref.~\cite{parker12}.
The resolution of our detector also compares well with that of 
other imaging devices with TOF capability that have been 
or are being developed for use at pulsed neutron sources.

Table~\ref{tbl:dcomp} shows a comparison of several such detectors,
including the RPMT~\cite{hirota05}, consisting of a 0.25-mm thick ZnS(Ag)/$^6$LiF 
scintillator coupled with a position sensitive photomultiplier tube,
a stacked-GEM (gas electron multiplier) detector~\cite{uno12,uno12_2} incorporating
a cathode and GEMs coated with a thin boron film and an FPGA readout system,
and a detector based on a boron-impregnated micro-channel plate (MCP) and 
Timepix readout chip~\cite{tremsin09,tremsin11}.
For the MCP, values for two operating modes are listed: a high-rate mode with 
spatial resolution dictated by the pixel size of the Timepix, and
a high-resolution mode using the charge-centroid method to improve resolution
at the cost of rate performance.
The spatial resolution of our prototype exceeds that of both the RPMT and GEM 
detectors,
while also exceeding the rate performance and efficiency of the RPMT.
Our detector also has a much lower gamma sensitivity (by over three 
orders of magnitude) than the scintillator-based RPMT.
No values are given for the gamma sensitivity of the remaining detectors,
but they are also expected to outperform the RPMT.
In particular, although the GEM-based detector should have gamma rejection 
similar to that of the $\mu$PIC,
the detailed tracking with TOT information should give our 
$\mu$PIC detector the advantage.
On the other hand, the stacked-GEM and MCP detectors show superior rate
performance and detection efficiency.
For the detection efficiency, however, that of our detector could be roughly 
doubled by adjusting the gas mixture to include more $^3$He, bringing it up to the
level of the stacked-GEM and MCP detectors.

\begin{table*}[ht]
\caption{\label{tbl:dcomp} Comparison of selected counting-type detectors
with neutron time-of-flight capability.
For the MCP detector, the values for an alternate high-resolution 
operating mode, when different from the normal mode,
are given in parentheses.
}
\begin{center}
\tabcolsep=0.08cm
\small
\begin{threeparttable}
\begin{tabular}{lccccccc}
\hline
Detector & Spatial     & Time       & Efficiency 
 & Counting rate & Gamma       & Typical & Ref. \\
         & resolution  & resolution & (at 25.3~meV)
 & (counts/s)    & Sensitivity &  area   & \\
\hline
$\mu$PIC & 100--120~$\mu$m & 0.6~$\mu$s & 18\% & 200~kcps\tnote{a} & $<10^{-12}$ 
 & $100$~cm$^2$ & \cite{parker12} \\
\hline
RPMT & 0.8~mm & 10~$\mu$s & $\sim$6.5\%\tnote{b} & $\sim$20~kcps & 10$^{-9}$
 & 60~cm$^2$ & \cite{hirota05} \\
stacked-GEM & 0.5~mm & $<0.1$~$\mu$s\tnote{c} & $\sim$25\%\tnote{b} & 12~Mcps
 & Not given & $100$~cm$^2$ & \cite{uno12,uno12_2} \\
MCP w/ Timepix & 55~$\mu$m (15~$\mu$m) & $\sim$1~$\mu$s & $\sim$50\% 
 & $>0.5$~Gcps ($\sim$kcps) & Not given & $4$~cm$^2$  & \cite{tremsin09,tremsin11} \\
\hline
\end{tabular}
\begin{tablenotes}
\item[a] \footnotesize{Counting rate will increase to the order of 
Mcps after encoder upgrade~\cite{parker12}}.
\item[b] \footnotesize{Estimated from quoted values (RPMT: 30\%@0.95~nm, 
GEM: 30\%@0.22~nm) using the appropriate neutron absorption cross-sections.}
\item[c] \footnotesize{Estimated from the electron drift velocity of 
Ar-CO$_2$ (70:30) gas and the given GEM spacing.}
\end{tablenotes}
\end{threeparttable}
\end{center}
\end{table*}

Future efforts to improve the spatial resolution of our detector 
will focus on optimization of the gas mixture and reduction of the 
gain variation and strip pitch of the $\mu$PIC.
As discussed in Ref.~\cite{parker12},
optimization of the gas mixture for reduced electron diffusion and/or
shorter track lengths is expected to 
provide a 10 to 15\% improvement in the spatial resolution, 
as determined by simulation.
Also, as noted in Sec.~\ref{sec:unif},
completely eliminating the gain variation would yield an improvement 
of around 15\%.
The gain variation may be reduced by improving the uniformity of the physical
structures of the $\mu$PIC through tighter manufacturing controls.
Furthermore, owing to the low operating gain,
it becomes possible to reduce the size of the pixel structures
in order to achieve a strip pitch as low as 200~$\mu$m,
allowing us to take fuller advantage of the optimized gas properties.
Using our GEANT4 simulation, we estimate that the spatial resolution
would be reduced by $\sim$45\% (or to less than 60~$\mu$m)
after making all of the above improvements/modifications to the detector,
hence achieving a spatial resolution similar to the MCP 
(in high-rate mode) and imaging plates.

Additionally,
the rate performance of our detector will be significantly improved
through a planned upgrade of the encoder hardware.
As shown in Ref.~\cite{parker12}, our current data acquisition system can 
handle raw data rates of $\sim$320~Mb/s (transferred to VME memory via a 32-bit
parallel bus at 10~MHz), 
corresponding to a neutron count rate of roughly 200~kcps (kilo counts per second)
for present detector conditions.
Preliminary testing of the new encoder system indicates an increase
of roughly a factor of four, 
resulting in a data rate of 12.8~Gb/s (or a neutron rate of 800~kcps).
For a given data transfer rate,
we can further increase the neutron count rate by
reducing the amount of data per event in two ways:
1) by increasing the stopping power of the gas for shorter 
proton-triton track lengths, and 
2) by performing some simple data processing on the FPGA.
For instance, by replacing Argon with Xenon to reduce track lengths 
from 8 to 5~mm and calculating TOT on the FPGA, the amount of data per 
neutron event would be reduced by a factor of 3.2, 
achieving a neutron rate of over 2.5~Mcps.
To approach the rate performance of the GEM-based detector,
we could further reduce the track length to less than 2~mm by increasing 
the gas pressure above 5~atm, resulting in another factor of 3 or 4 increase in
the neutron rate.
Unfortunately, this last step would also reduce the spatial resolution 
(to the order of the 0.4~mm strip pitch),
through a combination of increased scattering in the entrance window due to
increased vessel thickness,
increased electron diffusion,
and the fact that we could no longer perform detailed tracking with so 
few hits per event.
Such improvements to the spatial resolution and rate performance will be
the subject of future studies.

\section{Example: Resonance selective imaging}
\label{sec:watch}

Neutron resonance absorption techniques take advantage of the
tendency of some nuclides to preferentially absorb neutrons at specific energies,
unique to each nuclide, to study isotopic composition
and temperature~\cite{res:sato09}.
Resonance absorption can be directly observed by measuring the 
neutron transmission spectrum of the sample, calculated as:

\begin{equation}
\label{eq:trans}
Tr({\rm TOF}) = \frac{I({\rm TOF})}{I_{0}({\rm TOF})}
\end{equation}
where $I({\rm TOF})$ and $I_{0}({\rm TOF})$ are the intensities 
measured with and without the sample, respectively.
Due to the preferential absorption,
the transmission will show sharp drops near a resonance. 
Such resonances typically occur for neutron energies above 1~eV
and should appear in the TOF region between 0 and 1~ms for the 15-m flight path
used in our experiment.
The strength (depth) of the resonance dip gives the isotopic density,
while thermal broadening allows us to deduce the temperature within the sample.
The spatial distribution of the density and/or temperature can then be determined
by measuring the transmission point-by-point.
Using a 2-dimensional detector such as ours, this kind of measurement is
greatly simplified.

\begin{figure}[ht]
\centering
\includegraphics[width=7. cm,clip]{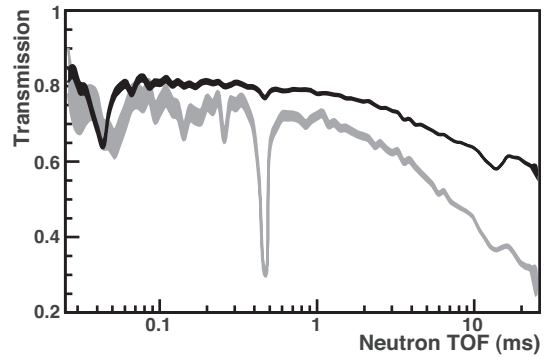}
\caption{\label{fig:wtrans} Neutron transmission spectra of a wristwatch.
The transmission versus TOF is plotted for the main body (black line)
and battery (gray line) of a wristwatch.
The thickness of the lines indicates $\pm 1\sigma$ statistical error.
Resonances due to copper and silver are visible around 0.04 and 0.46~ms,
respectively.
A Bragg edge is also visible around 15~ms.
}
\end{figure}

As an example, we considered an image of a wristwatch 
taken in February 2011 at NOBORU, under conditions
similar to those described in Sec.~\ref{sec:res}.
Transmission spectra, calculated via Eq.~\ref{eq:trans}, 
are shown in Fig.~\ref{fig:wtrans} for the
main body of the watch (black line) and the watch battery only (gray line).
The large resonance dip visible in the transmission of the watch battery
around 0.46~ms (or a neutron energy of $E_n = 5.2$~eV) is due to silver present in 
the battery's cathode
(along with many smaller dips at shorter TOFs).
A dip due to copper is also visible around 45~$\mu$s (or 580~eV)
in the main-body spectrum.

\begin{figure}[ht]
\centering
\includegraphics[width=7. cm,clip]{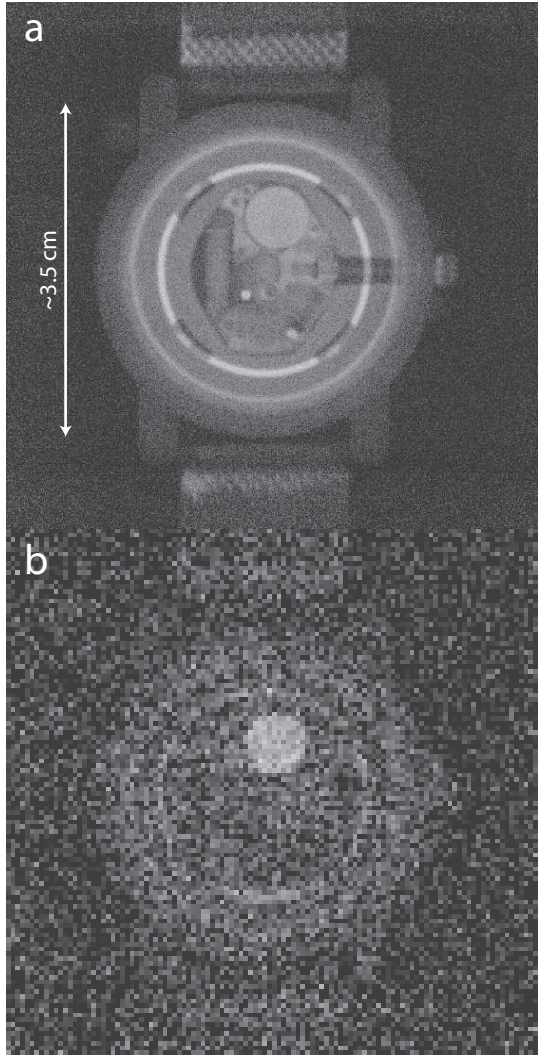}
\caption{\label{fig:wimg} Images of a wristwatch taken at NOBORU, J-PARC.
Radiographic images of a wristwatch are shown for neutron TOF ranges of
(a) ${\rm TOF} > 2.1$~ms and (b) $0.41 < {\rm TOF} < 0.51$~ms
(darker areas correspond to higher neutron transmission).
The TOF range for (b) corresponds to a resonance in the absorption cross-section
of silver at a neutron energy of 5.2~eV.
Bin sizes are (a) 80~$\mu$m and (b) 320~$\mu$m.
Light, horizontal bands at the top and bottom of (a) are due to the 
aluminum-backed tape used to hold the watch in place.
}
\end{figure}

Figs.~\ref{fig:wimg}(a) and \ref{fig:wimg}(b) show radiographic images
for neutrons with TOF $> 2.1$~ms and $0.41 < {\rm TOF} < 0.51$~ms, respectively.
(The gamma rejection cuts described in Sec.~\ref{sec:res} are applied in
both cases.)
The TOF range of \ref{fig:wimg}(b) selects neutrons with energy near 
the larger silver resonance, and
although statistics are greatly reduced 
(necessitating a larger bin size for the image),
the battery is clearly enhanced relative to the rest of the watch.
A quantitative study of the determination of nuclide density via
resonance absorption with a $\mu$PIC-based neutron imaging detector can 
be found in Ref.~\cite{harada12}. 

Finally, note that a Bragg edge is visible in the spectra
of Fig.~\ref{fig:wtrans} as a jump in the transmission around 15~ms. 
Such Bragg edges appear at the point where the neutron wavelength exceeds
the Bragg-scattering condition for a particular set of crystal planes,
and are observed for energies down to a few meV
(or for a TOF up to about 20~ms in the present case).
The Bragg edge observed here is most likely due to the metal body of the watch.
From the position of the Bragg edge, one can determine the crystal spacing,
and by studying the shape of the spectrum in the vicinity of the edge,
one can learn about crystal properties such as grain size and 
texture~\cite{bragg:sato09}.
Thus, in a single pulsed-beam measurement using a single detector,
it is possible to study both the 
isotopic composition and crystal structure of a sample in two dimensions.

\section{Conclusion}

We have introduced a neutron position reconstruction method based on 
template fitting that improved both the spatial resolution and 
event reconstruction efficiency of our $\mu$PIC-based, time-resolved 
neutron imaging detector.
The spatial resolution achieved with this method was found to be
approximately Gaussian with a sigma of $108.24 \pm 0.74$~$\mu$m
and a non-uniformity of 6.4\% RMSD
(reconstruction efficiency $\varepsilon_{rec} = 31.8$\%).
This represents a factor of three improvement over the resolution determined 
using the simple algorithm of our previous work~\cite{parker12}.
By using less restrictive cut conditions,
the reconstruction efficiency was greatly improved,
while worsening the spatial resolution by only 14\%
(consistency cut, $\varepsilon_{rec} = 66$\%) and 19\%
(TOF and gamma cuts only, $\varepsilon_{rec} = 100$\%).
Additionally,
after reducing the gain to the optimum value of 250 and correcting for beam
divergence,
the spatial resolution of the $\mu$PIC would improve to just under
100~$\mu$m for the tightest cut condition
(or $\sim$120~$\mu$m for TOF and gamma cuts only).
The optimum gain setting of 250 also corresponds to that shown in 
Ref.~\cite{parker12} to give an effective gamma sensitivity of 10$^{-12}$ or less.
Future improvements in the spatial resolution,
to better than 60~$\mu$m, may be possible
through optimization of the gas mixture for reduced electron diffusion, 
reduction of the gain variation of the $\mu$PIC 
through improved manufacturing controls,
and reduction of the strip pitch.

We also demonstrated in the example measurement of Sec.~\ref{sec:watch} that
by combining high-resolution imaging with event-by-event TOF measurement,
our detector becomes a powerful and flexible tool for radiographic studies
at pulsed neutron sources.
Using such a detector, we can simultaneously record the radiographic image, 
resonance absorption, and Bragg-edge transmission information point-by-point,
allowing a variety of material properties,
including nuclide composition and density, internal temperature, 
crystal spacing, grain size and texture, internal strain, etc., 
to be studied in a single measurement. 
By utilizing computed tomography techniques, such measurements can also be
extended to three dimensions,
allowing non-destructive study of various bulk properties.

\section*{Acknowledgements}

This work was supported by the Quantum Beam Technology Program of the Japan 
Ministry of Education, Culture, Sports, Science and Technology (MEXT).
The neutron experiments were performed at NOBORU (BL10) of the J-PARC/MLF
with the approval of the Japan Atomic Energy Agency (JAEA), Proposal No.~2009A0083.
The authors would like to thank the staff at J-PARC and the 
Materials and Life Science Experimental Facility 
for their support during our test experiments.
The authors would also like to thank M.~Ohi for providing the wristwatch 
used in our resonance absorption measurement.

%\section*{References}

%%
%%%%%%%%%%% end of main text %%%%%%%%%%%%%%%%%%

%% The Appendices part is started with the command \appendix;
%% appendix sections are then done as normal sections
%% \appendix

%% \section{}
%% \label{}

%% References
%%
%% Following citation commands can be used in the body text:
%% Usage of \cite is as follows:
%%   \cite{key}          ==>>  [#]
%%   \cite[chap. 2]{key} ==>>  [#, chap. 2]
%%   \citet{key}         ==>>  Author [#]

%% References with bibTeX database:

\bibliographystyle{elsarticle/model1-num-names}
%\bibliography{references.bib}
%\end{document}

%% Authors are advised to submit their bibtex database files. They are
%% requested to list a bibtex style file in the manuscript if they do
%% not want to use model1-num-names.bst.

%% References without bibTeX database:

\end{document}